\documentclass[letterpaper,journal]{IEEEtran}
\usepackage{amsmath,amssymb,amsfonts}
\usepackage{algorithmic}
\usepackage{algorithm}
\usepackage{array}
\usepackage[caption=false,font=normalsize,labelfont=sf,textfont=sf]{subfig}
\usepackage{cite}
\usepackage{graphicx}
\usepackage{multirow}
\usepackage{stfloats}
\usepackage{textcomp}
\usepackage{url}
\usepackage{url}
\usepackage{verbatim}
\usepackage{xcolor}

\begin{document}

\title{Scheduling for TWDM-EPON-Based Fronthaul Without a Dedicated Registration Wavelength}
\author{Akash~Kumar, Sourav~Dutta, and~Goutam~Das

\thanks{A. Kumar and G. Das are with the Department of Electronics and Electrical Communication Engineering, Indian Institute of Technology Kharagpur, Kharagpur 721302, India (e-mail: akash527845@kgpian.iitkgp.ac.in; gdas@ece.iitkgp.ac.in).} % <-this % stops a space
\thanks{S. Dutta is with the School of Computing and Electrical Engineering, Indian Institute of Technology Mandi, Mandi 175005, India (e-mail: sourav@iitmandi.ac.in).}% <-this % stops a space
}




\maketitle

\begin{abstract}
Centralized Radio Access Network C-RAN architecture imposes massive bandwidth demands on fronthaul links, along with stringent delay and jitter constraints. While Passive Optical Networks (PONs) offer a cost-effective solution, their standard registration mechanisms impose a critical bottleneck: the periodic registration windows required for Radio Unit (RU) discovery halt all upstream traffic on the associated channel, directly conflicting with fronthaul latency and jitter requirements. The ITU-T acknowledges this issue and recommends a dedicated wavelength for registration, but this comes at the cost of inefficient bandwidth utilization. Rather than adopting the same approach in Ethernet PON (EPON)-based fronthaul, we present an alternate solution wherein registration is carried out on one wavelength while active traffic is redistributed to the others. To this end, we propose a scheduling framework for Time and Wavelength Division Multiplexed (TWDM) EPON-based fronthaul that enables periodic registration without wasting an entire wavelength channel. This strategy ensures that the full wavelength pool is utilized for data transmission outside registration periods. Performance evaluation shows that the proposed method achieves up to a 71\% increase in the number of RUs supported for a given number of wavelength channels compared to the baseline scheme employing a dedicated registration wavelength.
\end{abstract}


\begin{figure*}[t]
    \centering
    \includegraphics[width=0.70\linewidth]{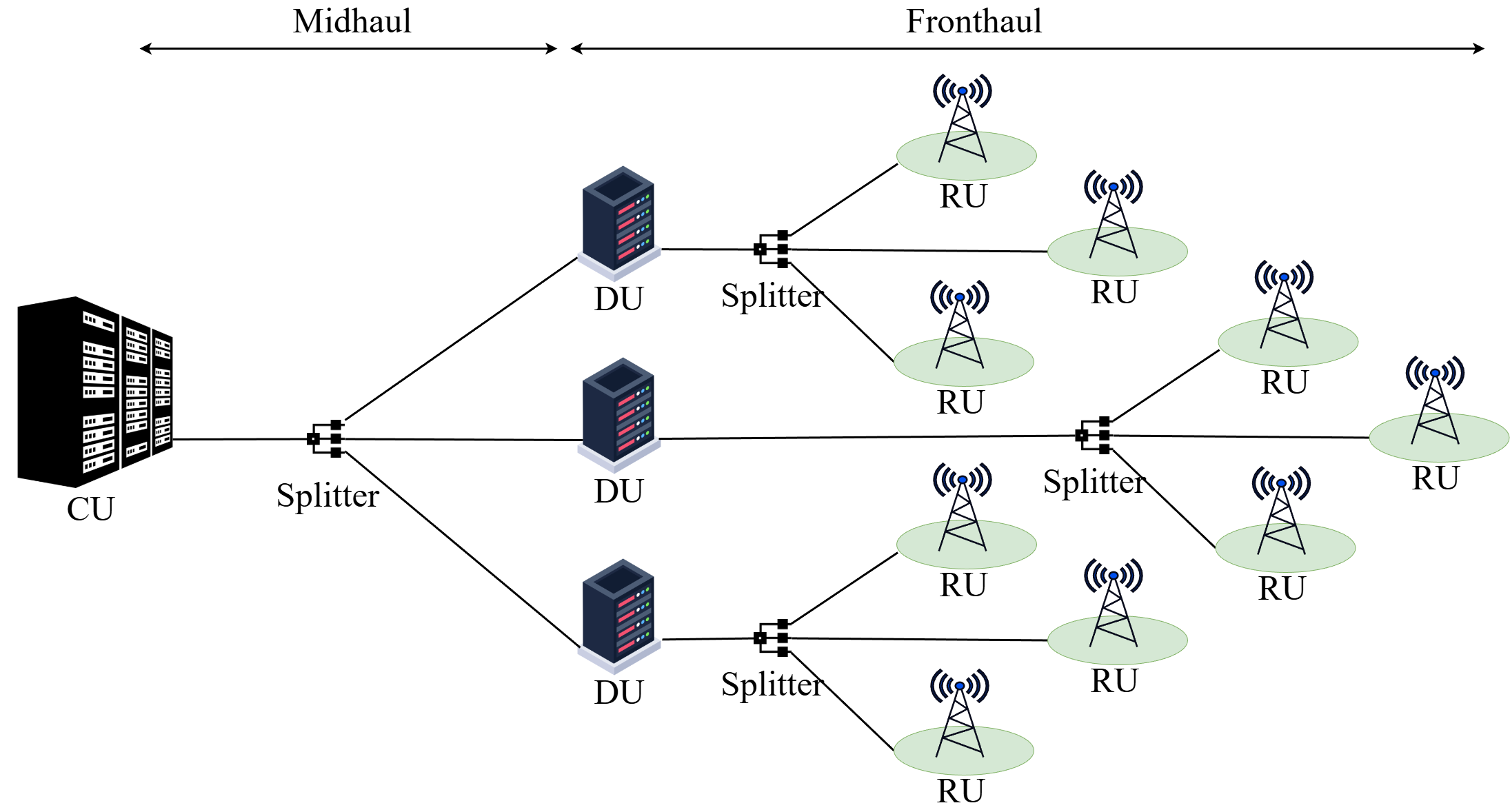}
    \caption{X-haul architecture illustrating the connectivity among the Central Unit (CU), Distributed Unit (DU), and Radio Units (RUs).}
    \label{fig:xhaul}
\end{figure*}

\section{Introduction}
\label{sec:intro}

Emerging applications such as multi-sensory extended reality (XR), autonomous driving, and industrial control systems demand unprecedented bandwidth, ultra-low end-to-end latency, and support for highly bursty traffic~\cite{6g, ITU2030}. The rapid growth of these applications exposes the last-mile access network as a critical performance bottleneck~\cite{6GRAN}. Centralized Radio Access Network (\mbox{C-RAN}) architectures, originally proposed to address the economic inefficiencies and limited coordination capabilities of traditional Distributed RAN (D-RAN) deployments, have therefore become increasingly critical~\cite{6GRAN, cmcccran}. Unlike D-RAN architectures, where baseband processing is fully localized at individual cell sites, \mbox{C-RAN} decouples baseband processing from remote radio hardware and consolidates part or all of these functions into centralized processing pools, thus enabling advanced inter-cell coordination and flexible sharing of baseband resources~\cite{cmcccran, cranoran, checkosurvey}.

To maximize centralization and interoperability, the \mbox{O-RAN} Alliance~\cite{oran-arch} standardizes a disaggregated architecture composed of Central Units (CUs), Distributed Units (DUs), and Radio Units (RUs), as shown in Fig.~\ref{fig:xhaul}. The partitioning of baseband processing across these units follows standardized functional splits defined in~\cite{3gpp-tr-38.801}. Specifically, a Higher Layer Split (HLS) determines the functional division between the CU and DU, while a Lower Layer Split (LLS) determines the division between the DU and RU.

Among the LLS options, Option~7 (sub-options~7-1, 7-2, and~7-3; intra-PHY functional splits) and Option~8 are of particular interest in this work. Option~7 partitions physical-layer processing between the DU and RU, whereas Option~8 centralizes the entire physical layer at the DU, leaving the RU responsible only for radio-frequency processing and over-the-air transceiving~\cite{3gpp-tr-38.801}. Under these options, the DU–RU link, referred to as the fronthaul, carries partially processed baseband data corresponding to different points in the PHY-layer processing chain for Option~7, and digitized time-domain radio samples for Option~8. Consequently, these options impose the most stringent fronthaul requirements, including extremely high bandwidth demands, sub-millisecond one-way delay budgets, and near-zero jitter tolerances~\cite{3gpp-tr-38.801, ecpri_spec}.

To meet these requirements, optical transport is generally preferred over wireless solutions (e.g., microwave and millimeter-wave (mmWave)), as the latter are fundamentally constrained by restricted spectrum, high interference, and limited capacity~\cite{pon-fronthaul}. Among optical access solutions, multi-wavelength Ethernet Passive Optical Networks (EPONs)\footnote{EPON employs a point-to-multipoint topology, connecting a central Optical Line Terminal (OLT) to multiple Optical Network Units (ONUs) via a passive optical splitter (see Fig.~\ref{fig:epon}).} have attracted significant interest due to their mature ecosystem, cost efficiency, and compatibility with deployed fiber infrastructure~\cite{multi_wave}. Within such fronthaul architectures, radio information is typically transported using the Common Public Radio Interface (CPRI)~\cite{cpri-spec} or its packet-based evolution, the enhanced CPRI (eCPRI) protocol~\cite{ecpri_spec}.

During real-time operation, network operators need to switch RUs on and off in order to adapt to diurnal traffic variations, reduce energy consumption, and enable network expansion~\cite{ITU-L1390}. Consequently, the fronthaul system requires a \emph{periodic} registration mechanism to discover and integrate newly activated RUs during normal operation (and not only at initial provisioning). To enable this, the Ethernet standard IEEE~802.3~\cite{reg_doc} inserts periodic \emph{registration windows} (referred to as \emph{Discovery windows} in Ethernet standards), during which all scheduled upstream transmissions are halted and the channel is used solely for registration. These windows must be at least as long as the round-trip time (RTT) of the farthest RU for upstream ranging and timing alignment. Such pauses in data transmission inevitably violate the stringent sub-millisecond delay and near-zero jitter requirements of fronthaul systems. This issue is examined in detail in Section~\ref{sec:motivation-and-background}.

This limitation is also present in ITU-T defined passive optical networks (such as G-PON, XG-PON, and NG-PON2), which similarly rely on 'quiet windows' for the registration process. To address this limitation in multi-wavelength architectures, the ITU-T recommends dedicating a wavelength channel exclusively to the registration process, while regular fronthaul traffic is carried over the remaining wavelengths~\cite{ITU-doc}. By isolating registration traffic from normal data transmission, this approach effectively eliminates the delay and jitter violations introduced by periodic registration windows. However, it does so at the cost of reserving a full wavelength channel for a control-plane function that requires only a minimal bandwidth. As a result, a substantial fraction of the available optical capacity remains underutilized.

Rather than adopting this bandwidth-inefficient ITU-T strategy in an EPON-based fronthaul, we aim for a solution that preserves spectral efficiency. This can be achieved by redistributing the traffic of active RUs across the remaining wavelengths during registration intervals. Building on this principle, this paper proposes a scheduling framework for a Time and Wavelength Division Multiplexed (TWDM) EPON-based fronthaul. In this scheme, the registration process is hosted on a selected wavelength, and traffic from active RUs is temporarily carried over the remaining ones. Outside registration periods, all the wavelengths remain available for fronthaul traffic, thereby satisfying fronthaul requirements without sacrificing bandwidth.

However, during registration periods, occupying one wavelength for registration temporarily reduces the available bandwidth, causing eCPRI traffic to accumulate in buffers and introduce additional delay. This excess delay must be controlled in order to satisfy fronthaul delay requirements. To this end, we formulate a delay-bounded optimization problem that maximizes the number of supported RUs for a given number of wavelength channels. Numerical results demonstrate that the proposed framework increases the number of supported RUs by up to 71\% compared to the ITU-T baseline approach of dedicating a wavelength channel for registration.

To summarize, our key contributions are as follows:
\begin{enumerate}
    \item We propose a traffic redistribution mechanism for multi-wavelength EPON fronthaul that enables periodic ONU registration on a selected wavelength while maintaining uninterrupted fronthaul traffic on the remaining ones.
    \item Building upon this, we develop a scheduling framework for TWDM-EPON-based fronthaul. We formulate a delay-constrained optimization problem to maximize the number of supported RUs and evaluate its performance against the dedicated-wavelength approach.
\end{enumerate}

The remainder of this paper is organized as follows. Section~\ref{sec:related-works} reviews prior work on fronthaul scheduling and ONU registration. Section~\ref{sec:motivation-and-background} provides the necessary background and outlines the motivation for this study. Section~\ref{sec:system-model-and-proposed-solution} presents the system model and the proposed traffic redistribution mechanism. Section~\ref{sec:optimization-formulation} presents the optimization problem formulation and solution methodology. Section~\ref{sec:results} presents the performance evaluation under varying system parameters. Section~\ref{sec:conclusion} concludes the paper.

\section{Related Works}
\label{sec:related-works}

A substantial portion of existing literature focuses primarily on efficient fronthaul scheduling and resource management, while largely omitting the challenges associated with RU registration. In particular,~\cite{jlt2025, hrsl, AtriSirPaper, active-wav, vbs-vpon, game-paper, mobilepon} investigate radio and fronthaul resource allocation using optimization-based frameworks to handle bursty traffic and control end-to-end latency. Further, some recent studies~\cite{mobilepon_ex, app, dl-dba, ewong} apply Machine Learning (ML) techniques to improve throughput by predicting real-time traffic patterns. However, these works~\cite{jlt2025, hrsl, AtriSirPaper, active-wav, vbs-vpon, game-paper, mobilepon, mobilepon_ex, app, dl-dba, ewong}, implicitly or explicitly, assume that RU registration follows standard procedures, thereby overlooking its impact on fronthaul traffic. Since standard EPON registration requires suspending upstream transmission, these approaches are applicable only when a dedicated registration wavelength is used.

To address discovery-induced violations without dedicating a wavelength, Uzawa \emph{et al.}~\cite{low_bw_reg} propose an uplink scheduling scheme based on network slicing for mobile fronthaul, Internet of Things (IoT), and control traffic. This approach assumes a MAC--PHY functional split (Option~6), where fronthaul traffic is packetized and inherently bursty. To carry out registration, the scheme inserts a discovery window within Transmission Time Interval (TTI) cycles with sufficient unallocated bandwidth. However, such idle intervals are unavailable for the splits considered in this work (Options~7 and~8), where mobile fronthaul traffic arrives continuously at the ONU throughout the entire TTI. Consequently, such a scheme is not applicable when employing the considered splits.

Attempting to reduce the registration window size, Kim \emph{et al.}~\cite{adaptive-tdm} proposed using Optical Time Domain Reflectometry (OTDR) to manually pre-assign ONUs to distance zones, and perform registration for each zone separately. However, such manual ranging in PONs is inherently unreliable due to human errors, fiber faults, signal superposition, and protection switching events. These errors risk incorrect zone assignments, causing registration requests to drift outside their allocated windows and collide with active traffic. Further, adaptive registration scheme for TWDM-PON in~\cite{adaptive-twdm} resizes the window based on the number of contending ONUs, but remains constrained by the minimum RTT-sized window requirement. For fronthaul deployments, this still implies interruption durations exceeding strict mobile latency budgets.

To carry out registration without halting upstream traffic, \emph{hitless} activation techniques~\cite{whisper, dsm} superimpose low-power discovery signals directly onto the active upstream channel. These overlay signals convey registration messages at power levels sufficiently low to ensure negligible interference with active services, yet strong enough to be extracted via electrical-domain coherent detection or filtering. However, such methods require non-standard optical transceivers and sophisticated signal processing. As a result, they remain experimental and unsuitable for near-term deployment within the current cost-sensitive EPON ecosystem.

In summary, prior works (i) optimize fronthaul scheduling without considering registration-induced violations; (ii) attempt mitigation by reducing the window size; or (iii) depend on non-standardized transceivers and signal processing. To the best of our knowledge, no existing solution enables ONU registration in EPON-based fronthaul using standard hardware without compromising spectral efficiency. This gap motivates the framework proposed in this work.

\section{Background and Motivation}
\label{sec:motivation-and-background}

\begin{figure}[t]
    \centering
    \includegraphics[width=1\linewidth]{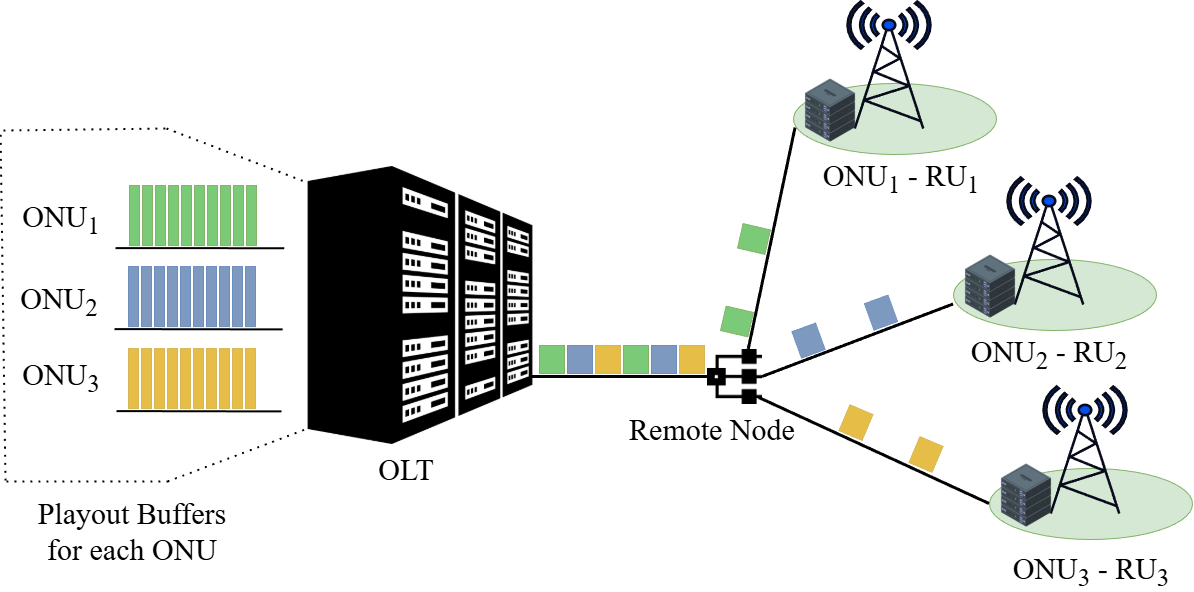}
    \caption{EPON-based fronthaul architecture illustrating the OLT, passive splitter, and multiple ONUs.}
    \label{fig:epon}
\end{figure}

In this section, we explain the fundamental limitation of the standard EPON registration procedure when applied to fronthaul systems, and highlight why it cannot be \emph{directly} adopted without modifications to upstream scheduling. We consider an EPON-based fronthaul network in which a centralized Optical Line Terminal (OLT) interfaces with the DU and aggregates upstream traffic from multiple Optical Network Units (ONUs), each serving a distinct RU (see Fig.~\ref{fig:epon}). The fronthaul traffic is transported using the eCPRI protocol, which requires high bandwidth and imposes strict delay and jitter constraints.

\subsection{Standard EPON Registration Procedure}

\begin{figure} [htbp]
    \centering
    \includegraphics[width=1\linewidth]{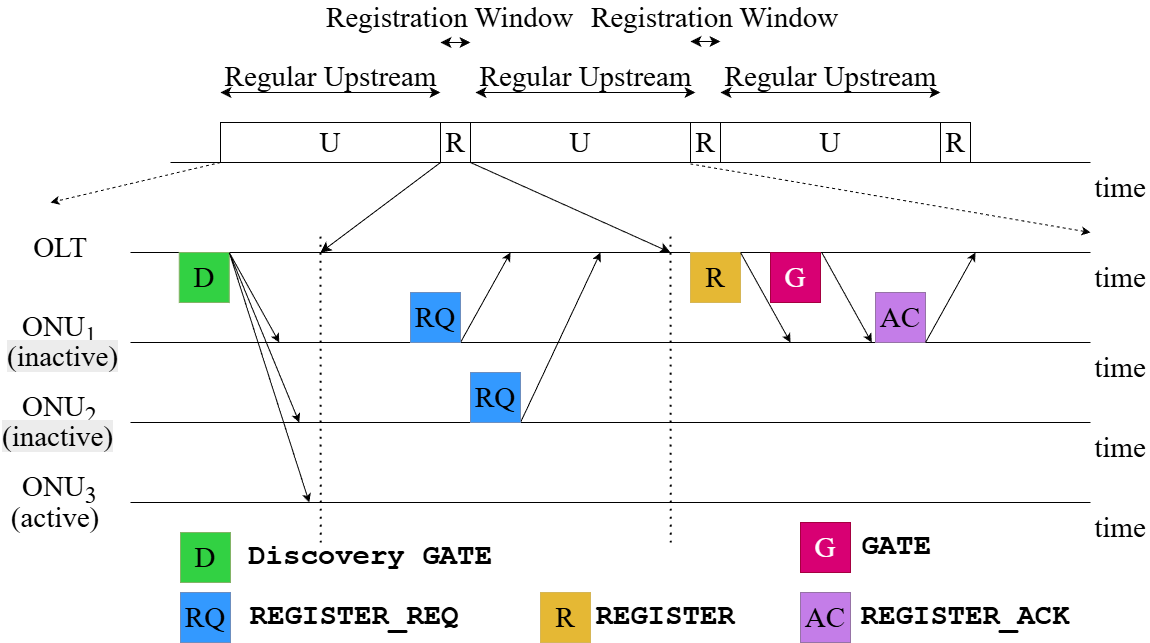}
    \caption{Standard EPON Registration Process.}
    \label{fig:reg-process}
\end{figure}

In standard EPON systems, the registration process follows IEEE~802.3 specifications~\cite{reg_doc} to admit newly connected or offline ONUs into the network. The upstream channel operation alternates between periods of normal upstream operation and scheduled registration windows (see Fig.~\ref{fig:reg-process}). During normal upstream operation, all registered ONUs transmit user data according to the grant schedule assigned by the OLT. Periodically, the OLT schedules a registration window, during which normal upstream transmissions are halted. This creates a "quiet window" where unregistered ONUs can request access.

The registration process follows a contention-based control message exchange between the OLT and unregistered ONUs. The OLT initiates the process by broadcasting a Discovery \texttt{GATE} message, which defines the start and length of the window. An unregistered ONU seeking to join the network waits until the designated start time and applies a random delay before sending a \texttt{REGISTER\_REQ} message. As multiple unregistered ONUs attempt to register in the same window, collisions may occur. In such events, the affected ONUs defer their registration attempts using a randomized back-off mechanism and retry in a subsequent window.

Upon successfully receiving a \texttt{REGISTER\_REQ}, the OLT assigns a Logical Link Identifier (LLID) to the ONU and responds with a \texttt{REGISTER} message carrying the required configuration parameters. Shortly thereafter, the OLT transmits a unicast \texttt{GATE} message to the newly registered ONU to establish its upstream transmission schedule. The ONU completes the handshake by sending a \texttt{REGISTER\_ACK} message, after which it can participate in normal upstream operation.

Although the registration procedure involves bidirectional control message exchange, a quiet window is required only in the upstream direction to accommodate contention-based \texttt{REGISTER\_REQ} messages. In contrast, all downstream control messages are centrally scheduled by the OLT and hence do not require a quiet window. Furthermore, the final upstream message, \texttt{REGISTER\_ACK}, is sent within an OLT-assigned grant, and thus also does not necessitate a quiet window. 

\subsection{Why standard EPON Registration is problematic}

To ensure that the ONU’s \texttt{REGISTER\_REQ} message is received within the scheduled window, the registration window duration must exceed the maximum round-trip time (RTT) between the OLT and the farthest ONU in the network. For typical fronthaul links of up to 20~km, this maximum RTT is approximately $200~\mu\text{s}$. To provide sufficient margin for random back-off and other delays, the registration window duration $T_{\mathrm{reg}}$ is typically configured slightly higher, around $250~\mu\text{s}$. 

\begin{figure}[htbp]
    \centering
    \includegraphics[width=0.80\linewidth]{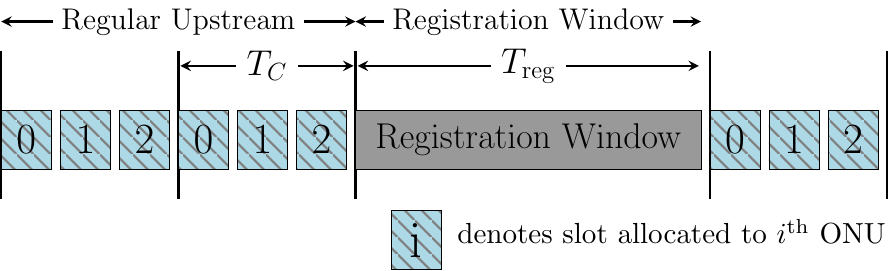}
    \caption{Conventional EPON upstream scheduling, highlighting the periodic interruption due to the mandatory registration window.}
    \label{fig:halted-traffic}
\end{figure}

This mandatory suspension of upstream traffic imposes a severe latency penalty. As shown in Fig.~\ref{fig:halted-traffic}, normal data transmission is blocked for the duration of $T_{\mathrm{reg}}$. In the worst-case scenario, where an eCPRI frame arrives immediately after its allocated slot, the frame is delayed by the full cycle duration $T_C$ plus the registration window $T_{\mathrm{reg}}$ before it can be served. Thus, the scheduling delay budget $D_b$ is lower-bounded by:
\begin{equation}
    D_b \geq T_{\mathrm{reg}} + T_C .
\end{equation}

In eCPRI-based fronthaul systems, the one-way delay budget for lower-layer split traffic is typically set below $250~\mu\text{s}$~\cite{3gpp-tr-38.801, ecpri_spec}. Given that the required registration window $T_{\mathrm{reg}}$ is also approximately $250~\mu\text{s}$, suspending upstream traffic for this duration inevitably violates the delay constraints, even without accounting for additional propagation or queuing overhead. As a result, the standard EPON registration procedure cannot be directly applied in fronthaul deployments without modifications to upstream scheduling.

As discussed in Section~\ref{sec:intro}, the ITU-T recommends dedicating a wavelength exclusively to the registration process to address this limitation~\cite{ITU-doc}. While this approach ensures compliance with fronthaul delay constraints, it does so at the cost of reduced spectral efficiency and underutilized bandwidth resources. In the next section, we present an alternative scheduling framework that enables periodic registration without violating fronthaul constraints, or wasting bandwidth.

\section{Proposed Framework}
\label{sec:system-model-and-proposed-solution}

We consider a TWDM-EPON fronthaul system in which each RU exchanges eCPRI traffic with the DU at a configured eCPRI traffic rate \(R_C\), although the instantaneous transmitted rate may be lower depending on the user traffic. This rate is assumed to be identical for all RUs so that we can focus on the core scheduling mechanism for clearer analytical exposition. Extensions to heterogeneous scenarios, where different RUs operate at different eCPRI rates, can be handled by the same framework by appropriately distributing unequal traffic loads across wavelengths to maintain balanced utilization. The remainder of the formulation remains unchanged.

We consider a fronthaul system that employs \(W\) wavelength channels in the upstream, each operating at \(R_E\) bits per second. Each wavelength channel is shared among \(N\) ONUs when the registration process is not being carried out. An ONU transmits only during its assigned time slot, resulting in a periodic but non-continuous transmission pattern. Based on this slotted structure, we denote each ONU as \( \mathcal{N}(\lambda, i_n) \), where \(\lambda \in \{0, 1, \ldots, W{-}1\}\) represents the wavelength index, and \(i_n \in \{0, 1, \ldots, N{-}1\}\) denotes its slot index in a cycle outside registration periods. This indexing is purely logical and does not imply a permanent physical binding of an ONU to a specific wavelength or slot.

In contrast to the slotted transmission described above, fronthaul traffic requires a continuous stream of packets at the receiver and is subject to stringent jitter constraints under the considered functional splits~\cite{3gpp-tr-38.801, ecpri_spec}. To address this, playout buffers are employed at the OLT (see Fig.~\ref{fig:epon}) to temporarily store and smooth the incoming data into a uniform output stream, thereby meeting the jitter requirement.

\begin{figure} [htbp]
    \centering
    \includegraphics[width=\linewidth]{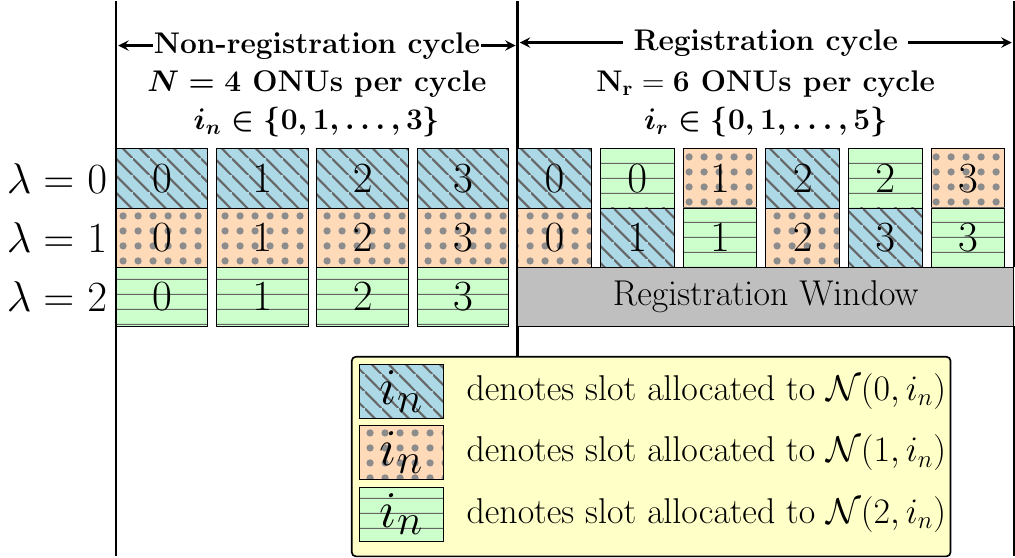}
    \caption{Proposed redistribution example for $N\!=\!4$ and $W\!=\!3$. During registration cycles, the remaining $W\!-1$ wavelengths support  $N_r\!=\!6$ ONUs per wavelength  (see Eq.~\ref{eq-nr-value}) .}
    \label{fig:onu-mapping}
\end{figure}

To enable the registration process, we place periodic registration windows on an arbitrary wavelength, while temporarily redistributing all the ONUs across the remaining \((W-1)\) wavelength channels during that interval, as shown in Fig.~\ref{fig:onu-mapping}. As bandwidth available for transmission is reduced in these cycles (referred to as Registration cycles here), extra data accumulates in the ONU buffers. In the cycles apart from the ones used for registration (referred to as Non-registration cycles here), all the available wavelengths are used for transmission, and the extra data accumulated during the registration cycles is cleared from the ONU buffers. The allocation follows a round-robin, row-wise order: we first assign the first slot on each wavelength to \(\mathcal{N}(0,0)\), \(\mathcal{N}(1,0)\), \(\ldots\), \(\mathcal{N}(W{-}2,0)\), then the next set of slots to \(\mathcal{N}(W{-}1,0)\), \(\mathcal{N}(0,1)\), \(\ldots\), \(\mathcal{N}(W{-}3,1)\), and so on, continuing until all \(\mathcal{N}(W{-}1,N{-}1)\) are reassigned (see Fig.~\ref{fig:onu-mapping}).

To maintain support for all the ONUs during registration cycles, the remaining \((W{-}1)\) wavelengths must collectively support the total \(N W\) ONUs. As a result, each of these wavelengths supports up to
\begin{equation}
    N_r = \left\lceil \frac{N \cdot W}{W - 1} \right\rceil
    \label{eq-nr-value}
\end{equation}
ONUs during registration cycles.

\begin{figure}[htbp]
    % \centering
    \includegraphics[width=0.9\linewidth]{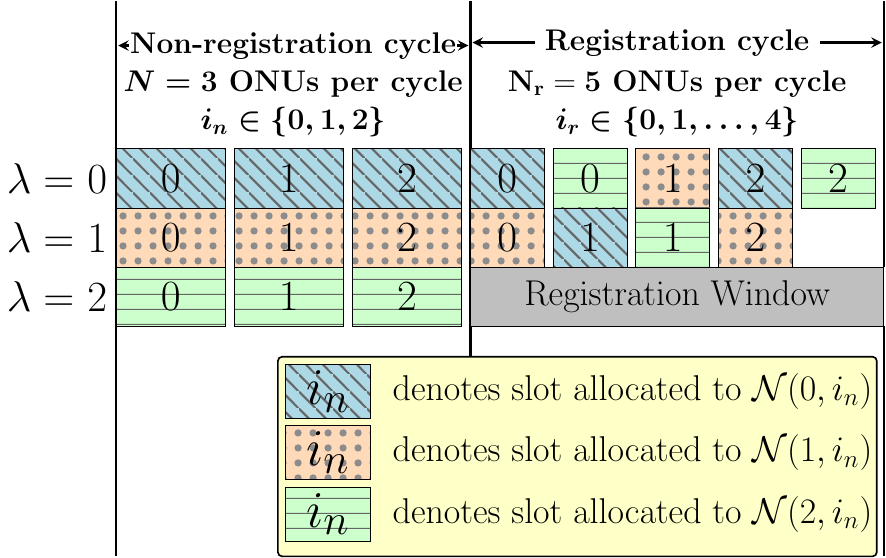}
    \caption{Example of ONU Redistribution for \(N = 3\) and \(W = 3\). The last slot of wavelength \(\lambda = 2\) remains vacant.}
    \label{fig:onu-mapping-2}
\end{figure}

Note that if \( N \) is not an exact factor of \( (W{-}1) \), some wavelengths will need to accommodate only \( (N_r{-}1) \) ONUs, whereas the remaining wavelengths will continue to support \( N_r \) ONUs each (see Fig.~\ref{fig:onu-mapping-2}). Since registration cycles last for a much smaller duration (on the order of a few hundred microseconds) compared to non-registration data cycles (on the order of a few hundred milliseconds), the presence of these vacant slots does not result in any significant throughput loss.

Under this redistribution, the slot index assigned to ONU \(\mathcal{N}(\lambda,i_n)\) during a registration cycle is given by
\begin{equation}
i_r = \left\lfloor \frac{W\, i_n + \lambda}{W - 1} \right\rfloor.
\label{eq:redis}
\end{equation}
where \(i_r \in \{0,1,\ldots,N_r{-}1\}\). The floor operation reflects the row-wise, round-robin assignment across the remaining \((W{-}1)\) wavelengths, mapping consecutive ONUs to successive slots until all \(N W\) ONUs are accommodated.

Note that the redistribution mechanism is independent of the wavelength hosting the registration process; hence, the registration window may be placed on any wavelength as required, either fixed or dynamically varied across cycles. Furthermore, the wavelength switching performed by ONUs to execute the proposed redistribution is only required at the transition points between registration and non-registration cycles, ensuring that at  least one full scheduling cycle is available for the wavelength switching.

As the system alternates between registration and non-registration cycles, ONUs experience non-uniform inter-slot gaps at cycle transitions. As a result, even within the same cycle, different ONUs may have different residual backlogs at the start of their assigned slots, inherited from the preceding cycle. In each granted slot, an ONU attempts to transmit and clear all buffered data within the allocated slot duration; if the slot is insufficient, the remaining backlog is deferred and served in subsequent slots.

The scheduling delay of an upstream frame is determined by the inter-slot gap and the residual backlog from the previous cycle. The playout buffer at the OLT absorbs these delay variations and aligns the output stream, such that all frames effectively experience the same maximum scheduling delay. Therefore, it suffices to compute the scheduling delays across all cycles and ensure that none exceeds the allowable delay budget. Based on this bound, we formulate, in the next section, a delay-constrained optimization problem to maximize the number of supported RUs.

\begin{figure*}[t]
    \centering
    \includegraphics[width=\textwidth]{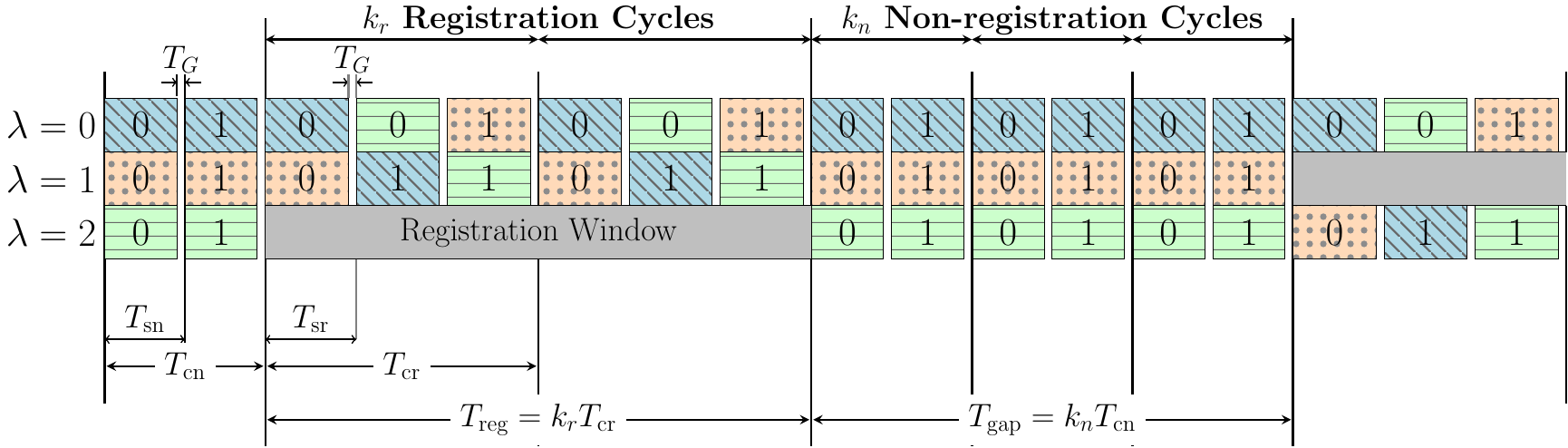}
    \caption{Example scheduling Diagram illustrating the alternation between $k_r=2$ registration cycles and $k_n=3$ non-registration cycles. The network configuration of $W=3$ upstream wavelengths and $N=2$ ONUs per wavelength (in non-registration cycles) implies a redistributed load of $N_r=3$ ONUs per active wavelength during the registration window.}
    \label{fig:scheduling-diagram}
\end{figure*}

\section{Optimization Problem Formulation}
\label{sec:optimization-formulation}

The primary objective of this formulation is to maximize the total number of supported RUs \(N W\), or equivalently, to maximize \(N\) for a given number of wavelengths \(W\). Since all RUs operate at an identical traffic rate $R_C$, maximizing the number of supported RUs is equivalent to maximizing bandwidth utilization. The system-level parameters, and decision variables used in the formulation are summarized in Table~\ref{tab:parameter-definitions}. 

The formulation begins by deriving the slot sizes required to serve the eCPRI traffic. Based on these, leftover backlog over successive cycles and the resulting worst-case scheduling delays are derived. Finally, the adopted solution methodology and experimental setup are detailed. Throughout the analysis, the traffic rate \(R_C\) is fixed to its maximum allowable value in order to capture the worst-case operating conditions. 

\begin{table}[htbp]
\centering
\renewcommand{\arraystretch}{1.2}
\caption{Definitions of System Parameters and Decision Variables}
\begin{tabular}{|c|l|}
\hline
\textbf{Symbol} & \textbf{Description} \\
\hline
$R_E$              & EPON line rate (per wavelength) \\
$R_C$              & Configured eCPRI traffic rate per ONU \\
$D_b$              & Scheduling delay budget \\
$T_{\mathrm{reg}}$ & Duration of a registration window \\
$T_{\mathrm{gap}}$ & Time gap between two registration windows \\
$G$              & Guard band duration \\
$\alpha$           & eCPRI frame size \\
$E_{\text{max}}$   & Maximum Ethernet payload size \\
$L_{\text{hdr}}$   & Ethernet header size \\
\hline
$W$                & Number of upstream wavelengths \\
$N / N_r$          & Number of ONUs per wavelength in a non-registration \\
                   & (non-reg.) / registration (reg.) cycle \\
$\mathcal{N}(\lambda, i_n)$
                   & ONU on wavelength $\lambda$ with non-reg. slot index $i_n$ \\
$i_r$              & Slot index of ONU $\mathcal{N}(\lambda, i_n)$ in a reg. cycle \\
\hline
$k_n / k_r$        & Number of consecutive non-reg. / reg. cycles \\
$f_n / f_r$        & eCPRI frames per slot in a non-reg. / reg. cycle \\
$p_n / p_r$        & Ethernet packets per slot in a non-reg. / reg. cycle \\
$T_{sn} / T_{sr}$  & Slot duration in a non-reg. / reg. cycle \\
$T_{cn} / T_{cr}$  & Duration of a non-reg. / reg. cycle \\
\hline
\(\Delta_{\mathrm{type}}^{(\lambda,i_n)}[k]\) 
                  & Time gap between start of ONU's $k$-th slot of specified \\
                  & type (reg./non-reg.) and start of preceding slot \\
\(B_{\mathrm{type}}^{(\lambda,i_n)}[k]\) 
                  & Time-equivalent leftover backlog after ONU's slot in \\
                  & $k$-th cycle of specified type (reg./non-reg.) \\
\(D_{\mathrm{type}}^{(\lambda,i_n)}[k]\) 
                  & Maximum scheduling delay for a frame in ONU's slot in \\
                  & $k$-th cycle of specified type (reg./non-reg.) \\
\hline
\end{tabular}
\label{tab:parameter-definitions}
\end{table}

\subsection{Slot sizes}

In a TWDM-PON system, each ONU is allocated fixed transmission slots. Under the assumption that all RUs generate traffic at the same eCPRI rate \(R_C\), these slots are identical for all ONUs within a given cycle type. However, to maintain generality, different slot sizes are considered for the two scheduling cycle types, namely registration and non-registration cycles, instead of enforcing a single common slot size. The corresponding slot durations are denoted by \(T_\mathrm{sr}\) and \(T_\mathrm{sn}\), respectively.

These slot sizes are determined by the maximum eCPRI payload (obtained as the product of the maximum number of eCPRI frames and the basic eCPRI frame size) that must be transmitted within a slot, plus the additional overhead introduced by the Ethernet packet headers. 

To ensure that a registration-cycle slot can serve \emph{up to} \(f_r\) eCPRI frames, the number of Ethernet packets required is
\begin{equation}
    p_r = \left\lceil \frac{f_r \alpha}{E_\mathrm{max}} \right\rceil,
    \label{eq-pr}
\end{equation}
where \(\alpha\) is the basic eCPRI frame size and \(E_\mathrm{max}\) is the maximum Ethernet payload size.

Since such a slot must be long enough to transmit \((f_r\alpha + p_r L_{\mathrm{hdr}})\) bits (where \(L_\mathrm{hdr}\) is the Ethernet header size), and must additionally include a guard band of duration \(G\) at the end of the slot, the registration-cycle slot size \(T_\mathrm{sr}\) must satisfy
\begin{equation}
    T_\mathrm{sr} \ge \frac{1}{R_E}\left(f_r \alpha + p_r L_{\mathrm{hdr}}\right) + G.
    \label{eq-Tsr}
\end{equation}
Substituting \(p_r\) from~\eqref{eq-pr}, the above constraint can be equivalently written as
\begin{equation}
    T_\mathrm{sr} \ge \frac{1}{R_E}\left(f_r \alpha + \left\lceil\frac{f_r\alpha}{E_\mathrm{max}}\right\rceil L_{\mathrm{hdr}}\right) + G.
    \label{eq-Tsr-exp}
\end{equation}

Similarly, to serve \emph{up to} \(f_n\) frames in a non-registration cycle slot, the corresponding packet count \(p_n\) and slot duration \(T_\mathrm{sn}\) satisfy
\begin{align}
    p_n &= \left\lceil \frac{f_n \alpha}{E_\mathrm{\max}} \right\rceil,  \label{eq-pn}\\
    T_\mathrm{sn} &\ge \frac{1}{R_E}\left(f_n \alpha + \left\lceil\frac{f_n\alpha}{E_\mathrm{\max}}\right\rceil L_{\mathrm{hdr}}\right) + G.     \label{eq-Tsn}
\end{align}

The non-linear ceiling operations in \eqref{eq-pr} and \eqref{eq-pn} are implemented via the equivalence
\begin{equation}
    p = \lceil X \rceil \iff X \leq p \leq X + 1 - \epsilon, \quad p \in \mathbb{Z},
\end{equation}
where $\epsilon$ is a sufficiently small positive constant.

Each registration cycle accommodates at most \(N_r\) ONUs while each non-registration cycle serves exactly \(N\) ONUs. Thus, the durations of the registration and non-registration cycles, denoted \(T_\mathrm{cr}\) and \(T_\mathrm{cn}\), respectively, are given by
\begin{subequations}
    \label{eq-tc}
    \begin{align}
    T_\mathrm{cr} &= N_r\,T_\mathrm{sr}, \label{eq-tcr} \\
    T_\mathrm{cn} &= N\,T_\mathrm{sn}. \label{eq-tcn}
    \end{align}
\end{subequations}

Now, for a cycle pattern if $k_r$ and $k_n$ consecutive registration and non-registration cycles, respectively, the resulting registration window size $T_{\mathrm{reg}}$ and inter-window gap $T_{\mathrm{gap}}$ are:
\begin{subequations}
\begin{align}
    T_{\mathrm{reg}} &= k_r T_\mathrm{cr} = k_r N_r T_\mathrm{sr}, \\
    T_{\mathrm{gap}} &= k_n T_\mathrm{cn} = k_n N T_\mathrm{sn}.
\end{align}
\end{subequations}

Figure~\ref{fig:scheduling-diagram} illustrates these parameters using an example with \(N = 2\) and \(W = 3\).

\subsection{Delay calculations}
\label{subsec:delay-calculations}

This subsection characterizes the worst-case scheduling delay experienced by an ONU over alternating registration and non-registration cycles. The scheduling delay varies across cycles due to differences in the inter-slot gaps; however, the playout buffer at the receiver absorbs these variations and reconstructs a uniform eCPRI stream toward the DU. Consequently, it suffices to compute the scheduling delays associated with each slot and ensure that all such delays remain below the given limit.

We first define some quantities for use throughout the delay calculations. For the ONU \(\mathcal{N}(\lambda,i_n)\):
\begin{itemize}
    \item \(D_{\mathrm{type}}^{(\lambda,i_n)}[k]\) denotes the maximum scheduling delay experienced by a frame transmitted in the ONU’s slot in the \(k\)-th cycle of the specified type. In later expressions, we use \(D_{\mathrm{reg}}^{(\lambda,i_n)}[k]\) and \(D_{\mathrm{nr}}^{(\lambda,i_n)}[k]\) to refer to registration and non-registration cycles, respectively.
    \item \(\Delta_{\mathrm{type}}^{(\lambda,i_n)}[k]\) denotes the time gap between the beginning of the ONU’s \(k\)-th slot of the specified type and the beginning of its immediately preceding slot.
    \item \(B_{\mathrm{type}}^{(\lambda,i_n)}[k]\) denotes the time-equivalent leftover backlog immediately after the ONU’s slot in the \(k\)-th cycle of the specified type.
\end{itemize}

The delay analysis follows a simple principle: for any given slot, the maximum scheduling delay of a frame equals the time elapsed since the ONU’s previous transmission opportunity plus any backlog remaining at the end of that previous slot. This relationship can be expressed in the generic form
\begin{equation}
    D_{\mathrm{type}}^{(\lambda,i_n)}[k]
    = \Delta_{\mathrm{type}}^{(\lambda,i_n)}[k]
      + B_{\mathrm{type}}^{(\lambda,i_n)}[k-1].
    \label{eq:delay_identity}
\end{equation}

The residual backlog evolves linearly across cycles as a function of the inter-slot gap and the amount of data transmitted in each slot. Specifically, the backlog after a given slot equals the backlog remaining after the previous slot, plus the data accumulated during the intervening gap, minus the data transmitted in the current slot.  This evolution is expressed as
\begin{equation}
    B_{\mathrm{type}}^{(\lambda,i_n)}[k]
    = B_{\mathrm{type}}^{(\lambda,i_n)}[k-1]
      + \Delta_{\mathrm{type}}^{(\lambda,i_n)}[k]
      - \frac{f_{\mathrm{type}}\alpha}{R_C},
    \label{eq:leftover_identity}
\end{equation}
Here, \(f_{\mathrm{type}} = f_r\) for registration cycles and \(f_{\mathrm{type}} = f_n\) for non-registration cycles.

It is important to note that equations~\eqref{eq:delay_identity} and \eqref{eq:leftover_identity} require special interpretation for the cycle index \(k = 0\). In this case, the terms with index \(k=-1\) refer to the immediately preceding scheduling cycle, which is the last cycle of the opposite type. 

\subsubsection{Gap between successive slots}

\begin{figure}[htpb]
    \centering
    \includegraphics[width=\linewidth]{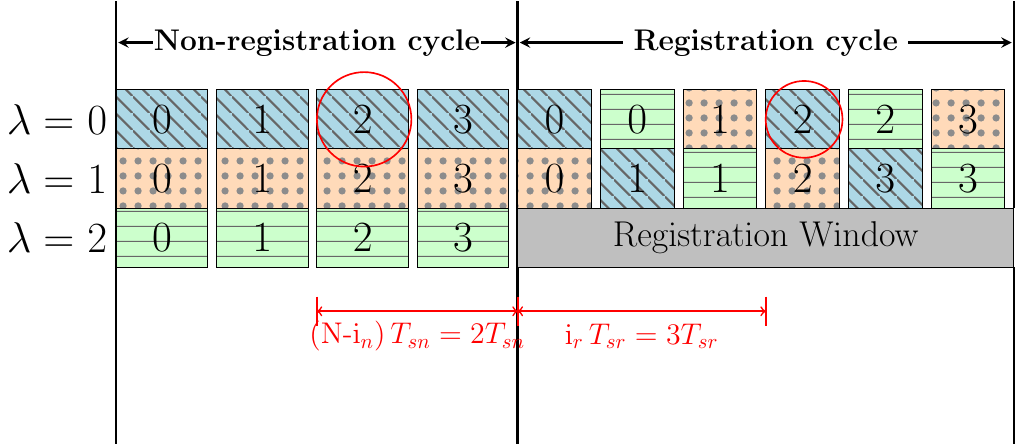}
    \caption{Illustration of Inter-slot Gap \(\Delta_{\mathrm{reg}}^{(\lambda,i_n)}[0]\) calculations with example of ONU \(\mathcal{N}(0,2)\). \(\Delta_{\mathrm{nr}}^{(\lambda,i_n)}[0]\) is obtained in an analogous manner.}
    \label{fig:gapreg0}
\end{figure}

Within a cycle, the inter-slot gap is fixed and equal to the cycle duration, i.e., \(T_{\mathrm{cn}}\) for non-registration cycles and \(T_{\mathrm{cr}}\) for registration cycles. A non-uniform inter-slot gap arises only at cycle transition points, where the gap depends on (i) the number of residual slots in the preceding cycle and (ii) the slot durations associated with registration and non-registration cycles. This calculation is illustrated in Fig.~\ref{fig:gapreg0}.
\begin{subequations}
\label{eq-gap}
\begin{align}
\Delta_{\mathrm{reg}}^{(\lambda,i_n)}[0]
    &= (N - i_n)\,T_\mathrm{sn} + i_r\,T_\mathrm{sr}, \label{eq-gapreg0} \\
\Delta_{\mathrm{reg}}^{(\lambda,i_n)}[k]
    &= T_\mathrm{cr}, \qquad k \in \{1,2,\dots,k_r-1\}, \label{eq-gapregk} \\
\text{For Non-registra} & \text{tion cycles}  \notag & \\
\Delta_{\mathrm{nr}}^{(\lambda,i_n)}[0]
    &= i_n\,T_\mathrm{sn} + (N_r - i_r)\,T_\mathrm{sr}, \label{eq-gapnonreg0} \\
\Delta_{\mathrm{nr}}^{(\lambda,i_n)}[k]
    &= T_\mathrm{cn}, \qquad k \in \{1,2,\dots,k_n-1\}.  \label{eq-gapnonregk} 
\end{align}
\end{subequations}

\subsubsection{Leftover data after each slot}

We evaluate the evolution of the leftover backlog in the ONU buffer by traversing the scheduling timeline forward, starting from the transition point where registration cycles begin. The leftover backlog at the end of the last non-registration cycle is zero, since non-registration slots are sized to serve all arriving traffic using the full set of \(W\) wavelengths. Consequently, backlog accumulation begins only with the onset of registration cycles.

\paragraph{First registration cycle}
During the gap of duration \(\Delta_{\mathrm{reg}}^{(\lambda,i_n)}[0]\), frames arrive at a maximum rate \(R_C\), while at most \(f_r\) frames can be transmitted in the subsequent slot. If the accumulated arrivals exceed the slot capacity, a positive backlog remains; otherwise, the backlog is zero. Accordingly,
\begin{equation}\label{eq:leftover_reg0}
    B_{\mathrm{reg}}^{(\lambda,i_n)}[0]
    =
    \max\left\{0,\;
    \Delta_{\mathrm{reg}}^{(\lambda,i_n)}[0] - \frac{f_r\alpha}{R_C}\right\}.
\end{equation}

The max-operator in \eqref{eq:leftover_reg0} is linearized using a standard Big-M formulation via the equivalence
\begin{equation}
    B = \max \{ 0, X \} \iff
    \begin{cases}
       X \leq M \delta, \\
       B \geq X, \\
       B \leq X + M(1 - \delta), \\
       B \leq M \delta, \\
       B \geq 0,
    \end{cases}
\end{equation}
where $M$ is a large positive constant and $\delta$ is a binary indicator variable defined uniquely for each ONU $\mathcal{N}(\lambda, i_n)$.

\paragraph{Subsequent registration cycles}

In a registration cycle, data accumulates during the inter-slot gap of duration \(T_\mathrm{cr}\), while at most \(f_r\) frames are transmitted in the corresponding slot. The difference between the accumulated and transmitted data determines the net backlog per registration cycle. Accordingly, the residual backlog increases linearly as
\begin{align}
    B_{\mathrm{reg}}^{(\lambda,i_n)}[k]
    = B_{\mathrm{reg}}^{(\lambda,i_n)}[0] + k\left(T_\mathrm{cr} - \frac{f_r\alpha}{R_C} \right),
    \label{eq:leftover_reg_k} \\
    \forall \;k \in \{0,1,\dots,k_r-1 \} \notag
\end{align}

\vspace{-0.4 em}
\paragraph{First non-registration cycle}

Using \eqref{eq:leftover_identity}, the leftover immediately after the first non-registration slot is
\begin{equation}
    B_{\mathrm{nr}}^{(\lambda,i_n)}[0]
    = B_{\mathrm{reg}}^{(\lambda,i_n)}[k_r-1]
      + \Delta_{\mathrm{nr}}^{(\lambda,i_n)}[0]
      - \frac{f_n\alpha}{R_C}. 
    \label{eq:leftover_nr_0}
\end{equation}

\paragraph{Subsequent non-registration cycles}

During the non-registration cycles, the transmission capacity of each slot exceeds the amount of data arriving during the corresponding cycle, resulting in a gradual reduction of the buffered backlog. Accordingly,
\begin{align}
    \label{eq:leftover_nr_k}
    B_{\mathrm{nr}}^{(\lambda,i_n)}[k]
    = B_{\mathrm{nr}}^{(\lambda,i_n)}[0]
      - k \left(\frac{f_n\alpha}{R_C} - T_\mathrm{cn} \right), \\
    \forall \; k \in \{0,1,\dots,k_n-1\}. \notag
\end{align}

This expression indicates that the residual backlog decreases monotonically over successive non-registration cycles, since the transmission capacity of each slot exceeds the data accumulated during the corresponding inter-slot gap. From \eqref{eq:delay_identity}, we can see that a reduction in backlog directly leads to a reduction in the associated scheduling delay. Consequently, the maximum scheduling delay is experienced either in the last non-registration or the first non-registration cycle, and it suffices to evaluate the delay only for these cycles. 

To ensure that the excess data accumulated during registration cycles is fully drained within the non-registration phase, we impose the constraint:
\begin{equation}
    B_{\mathrm{nr}}^{(\lambda,i_n)}[k_n-1] \le 0
    \label{eq:delta_nr_cleared}
\end{equation}
Although \(B_{\mathrm{nr}}^{(\lambda,i_n)}[k]\) represents a backlog quantity and therefore has no physical meaning when negative, allowing such values simplifies the formulation by avoiding explicit truncation without affecting the worst-case delay analysis.

\subsubsection{Maximum delay}

As explained earlier, the maximum scheduling delay occurs either in the final registration cycle or in the first non-registration cycle; therefore, it suffices to evaluate the scheduling delay expressions for these two cases. Substituting the inter-slot gaps in \eqref{eq-gap} and the residual backlog expressions in \eqref{eq:leftover_reg0}--\eqref{eq:leftover_nr_k} into the delay identity in \eqref{eq:delay_identity} yields the corresponding worst-case scheduling delays. Also, we recall that the terms indexed by \((k=-1)\) in \eqref{eq:delay_identity} and \eqref{eq:leftover_identity} correspond to the last scheduling cycle of the opposite type. 
\begin{subequations}
\label{eq-db}
\begin{align}
 D_{\mathrm{reg}}^{(\lambda,i_n)}[k_r-1]
        &= \Delta_{\mathrm{reg}}^{(\lambda,i_n)}[k_r-1] + B_{\mathrm{reg}}^{(\lambda,i_n)}[k_r-2] \le D_b \\
D_{\mathrm{nr}}^{(\lambda,i_n)}[0]
    &= \Delta_{\mathrm{reg}}^{(\lambda,i_n)}[0]+ B_{\mathrm{reg}}^{(\lambda,i_n)}[k_r-1]
\end{align}
\end{subequations}

\subsection{Solving the optimization problem}

The optimization problem selects the integer variables \(f_n\), \(f_r\), \(k_n\), and \(k_r\), along with the real-valued slot sizes \(T_\mathrm{sn}\) and \(T_\mathrm{sr}\), to maximize the number of  ONUs per wavelength \(N\), subject to the constraints \eqref{eq-Tsr-exp}, \eqref{eq-Tsn}, \eqref{eq:delta_nr_cleared}, and \eqref{eq-db}.

The optimization problem becomes difficult if \(N\), the number of ONUs per wavelength, is treated as a decision variable, since \(N\) directly determines the number of ONU-indexed variables and constraints in the formulation. In particular, the binary variables introduced to model the leftover backlog in \eqref{eq:leftover_reg0}, as well as the auxiliary variables used to represent the scheduling delay, inter-slot gaps, and leftover backlog, are defined on a per-ONU basis. Thus, allowing \(N\) to vary would result in an optimization problem with variable dimension due to variable number of variables and constraints. To address this difficulty, a search-based solution approach is adopted.

We start from the maximum feasible value of $N$ and iteratively decrease it until a valid solution satisfying all system constraints is found. This approach treats each iteration as a feasibility problem. The upper limit for $N$ is given by:
\begin{equation}
    N_{\text{max}} = \left\lfloor \frac{R_E}{R_C} \right\rfloor,
\end{equation}

In each iteration, \(N\) is treated as a constant. The resulting formulation still includes quadratic terms such as \(k_n f_n\) and \(k_r f_r\), making each iteration a Mixed Integer Quadratic Programming (MIQP) feasibility problem.

\begin{figure*}[t]
    \centering
    \setlength{\tabcolsep}{0pt} 
    \begin{tabular}{c c c}
        \includegraphics[width=0.33\textwidth]{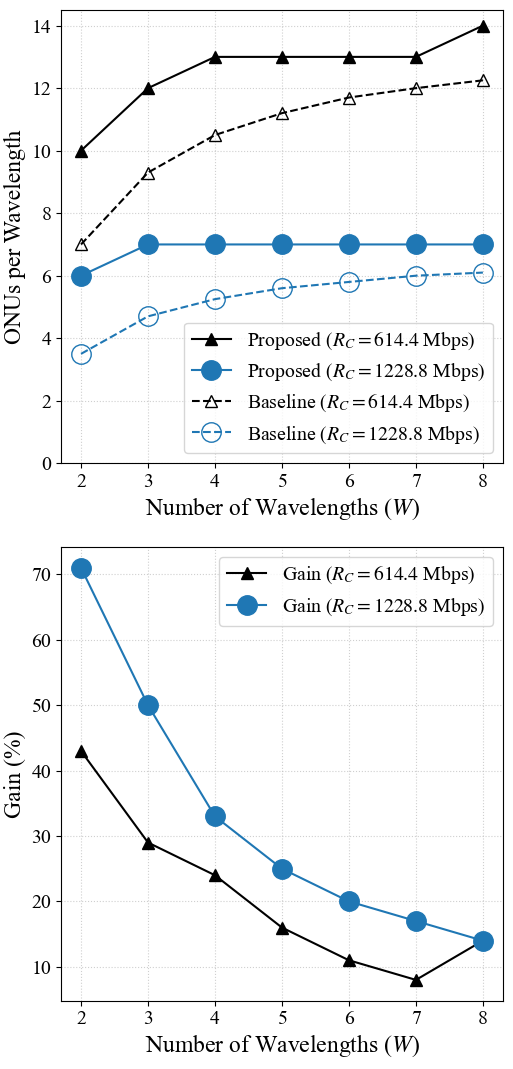} & 
        \includegraphics[width=0.327\textwidth]{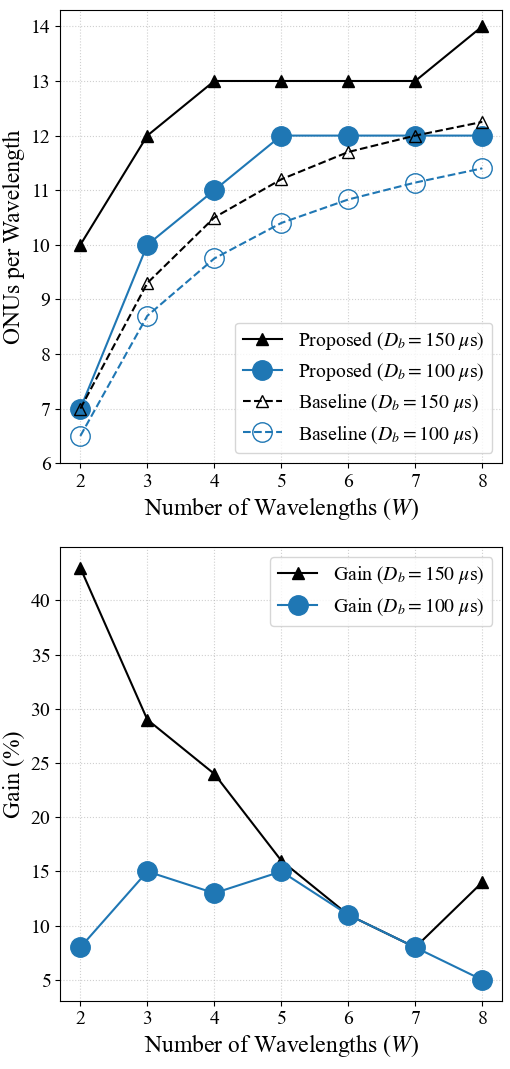} & 
        \includegraphics[width=0.33\textwidth]{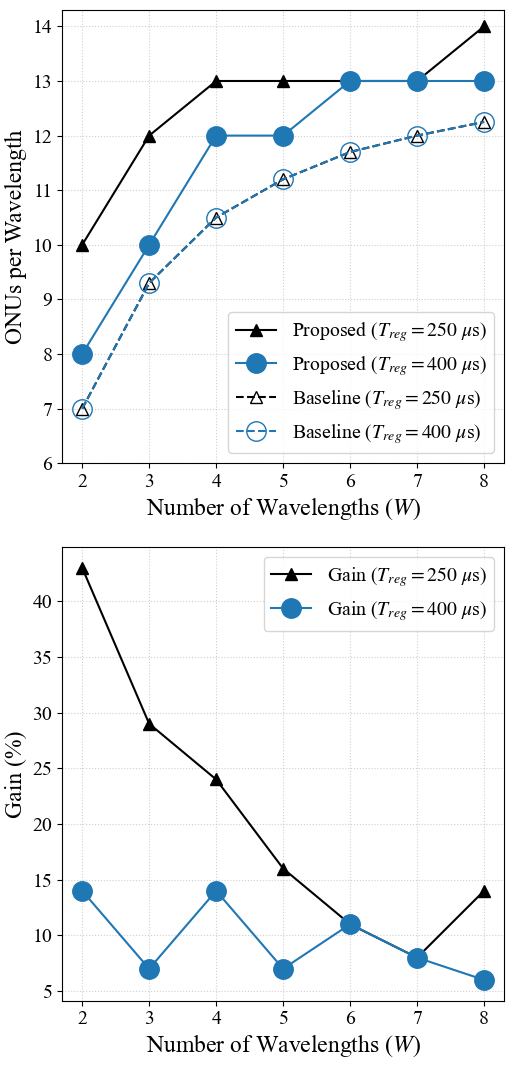} \\
        \parbox{0.33\textwidth}{\centering \footnotesize (a) Varying RU eCPRI traffic rate $R_C$} &
        \parbox{0.33\textwidth}{\centering \footnotesize (b) Varying Scheduling Delay Budget $D_b$} & 
        \parbox{0.33\textwidth}{\centering \footnotesize (c) Varying Registration Window size $T_{\mathrm{reg}}$} 
    \end{tabular}
    \caption{Performance comparison as a function of the number of wavelengths $W$ under varying system parameters. The top row illustrates the number of supported ONUs per wavelength, while the bottom row depicts the percentage gain achieved by the proposed framework.}
    \label{fig:plot}
\end{figure*}

\section{Results}
\label{sec:results}

In this section, we evaluate the performance of the proposed scheduling framework in terms of the total number of supported RUs for a given number of wavelengths \(W\).  The results are compiled in Fig.~\ref{fig:plot}, where we evaluate the performance gain achieved by the proposed framework for three scenarios: (i) varying eCPRI traffic rates \(R_C\) (ii) varying scheduling delay budget \(D_{\text{b}}\), and (iii) varying registration window size \(T_{\mathrm{reg}}\). The results are compared against a baseline approach in which one wavelength is reserved exclusively for registration,  as suggested in the ITU-T recommendation~\cite{ITU-doc}. The simulation parameters used, unless stated otherwise, are listed in Table~\ref{tab:parameters}.

The formulated optimization problem was solved using the Gurobi Optimizer~\cite{gurobi} on a machine equipped with an AMD Ryzen 7 5825U processor and 16~GB of RAM. Due to the small number of decision variables and the tight constraint set, each feasibility instance was solved within a few tens of milliseconds. Moreover, since \(N\)  and \(W\) were fixed in each iteration and the optimization only checked feasibility, the runtime per iteration remained largely independent of them. As a result, even with the iterative search required to determine \(N\), the total execution time remained computationally tractable and was typically below one second, even for large-scale scenarios involving hundreds of ONUs.

\begin{table}[htbp]
\centering
\renewcommand{\arraystretch}{1.2}
\caption{Simulation Parameters (unless stated otherwise)}
\begin{tabular}{|c|l|c|}
\hline
\textbf{Symbol} & \textbf{Description} & \textbf{Value} \\
\hline
$R_E$              & EPON line rate (per wavelength)        & 10 Gbps \\
$R_C$              & Configured eCPRI traffic rate per ONU  & 614.4 Mbps \\
$D_b$              & Scheduling delay budget                & 150~$\mu$s \\
$T_{\mathrm{reg}}$ & Duration of a registration window      & 250~$\mu$s \\
$T_{\mathrm{gap}}$ & Gap between two registration windows   & 100 ms \\
$G$              & Guard band duration                    & 1~$\mu$s \\
$\alpha$           & eCPRI frame size                       & 16 bytes \\
$E_{\text{max}}$   & Maximum Ethernet payload size          & 26 bytes \\
$L_{\text{hdr}}$   & Ethernet header size                   & 1500 bytes \\
\hline
\end{tabular}
\label{tab:parameters}
\end{table}

\subsection{Impact of Number of Wavelengths Used}
\label{sec:w-var}

For a given set of system parameters, suppose that the dedicated-wavelength baseline supports up to \(n\) RUs on each data-carrying wavelength. Since one wavelength is permanently reserved for registration, the effective system-wide capacity averages to \(n(W-1)/W\) RUs per wavelength. In contrast, the proposed framework utilizes all \(W\) wavelengths for fronthaul traffic. Although the number of supported RUs per wavelength may be slightly lower than \(n\) due to backlog accumulation during registration periods, the resulting aggregate capacity is consistently higher.  Thus, the relative gain is usually most pronounced for small values of \(W\), where the baseline suffers a substantial capacity penalty by dedicating a fraction \(1/W\) of its bandwidth to registration. As \(W\) increases, this overhead becomes less significant relative to the total system capacity, resulting in a gradual decrease in overall gain. 

However, the system behaves differently when constraints like the delay budget or registration window are tight. This is because at small $W$, the baseline has to cram all its traffic into just $W-1$ wavelengths, which spikes the load per channel. In such cases, that extra load becomes the main bottleneck, eating up the capacity and shrinking the gain. 

\subsection{Impact of eCPRI Traffic Rate}
\label{sec:rc-var}

In Fig.~\ref{fig:plot}(a), we examine the impact of the RU eCPRI traffic rate \(R_C\) on system performance, while keeping \(T_{\mathrm{reg}} = 250~\mu\text{s}\) and \(D_b = 150~\mu\text{s}\) fixed. It is observed that the proposed framework achieves a significantly higher relative gain as $R_C$ increases. This behavior follows from the analysis presented below.

The duration of a scheduling cycle (whether registration or non-registration) is determined by the number of supported ONUs per wavelength (see Eq.~\eqref{eq-tc}). Given a cycle duration $T$, servicing $N$ ONUs with a guard band $G$ yields an approximate upstream efficiency of $(T - NG)/T$. Thus, to meet the bandwidth requirements for $N$ ONUs, the system must satisfy:
\begin{equation}
    \frac{T - N G}{T} \geq \frac{N R_C}{R_E} \iff T \geq \frac{N G R_E}{R_E - N R_C}
\end{equation}
This inequality shows that a minimum cycle duration $T$ is required for a fixed set of system parameters $N, R_E, R_C, \text{and } G$, and a higher $N$ clearly necessitates a longer cycle duration $T$.

% While a lower $R_C$ accommodates a larger number of ONUs, the resulting linear increase in total guard band overhead ($NG$) necessitates a larger $T$ to satisfy the bandwidth requirements.

Referring back to Eq.~\eqref{eq:delay_identity}, the total delay experienced by a frame is determined by the cycle duration (the interval between transmission slots) and the backlog accumulated from previous cycles. This sum must remain within the strict scheduling delay budget. When $R_C$ is high (implying a smaller $N$), a shorter cycle duration $T$ suffices, leaving a generous margin within the delay budget to absorb the backlog from registration cycles. Conversely, at lower $R_C$, the increase in $N$ necessitates a larger cycle, and reduces the margin available for backlog. Consequently, delay constraints satisfied at a higher $R_C$ may be violated at a lower $R_C$, even if $N$ is scaled proportionally, compelling the system to support fewer ONUs. In contrast, the dedicated wavelength approach remains largely unaffected by these dynamics, as it doesn't have any special registration cycles where extra backlog is accumulated. This explains why the proposed scheme achieves a higher gain for higher $R_C$.

\subsection{Impact of Delay Budget}
\label{sec:delay-budget}

In Fig.~\ref{fig:plot}(b), we compare the gain achieved for two scheduling delay budgets, \(D_b = 100~\mu\text{s}\) and \(150~\mu\text{s}\), while keeping \(T_{\mathrm{reg}} = 150~\mu\text{s}\) and \(R_C = 614.4\)~Mbps fixed. The gain achieved by the proposed framework increases noticeably with a more relaxed delay budget. This trend arises because a tighter delay budget limits the backlog that can be accumulated during registration cycles, thereby constraining the number of supported RUs. In contrast, a relaxed delay budget enables larger backlog absorption and supports more RUs.

\subsection{Impact of Registration Window Size}
\label{sec:regwin}

In Fig.~\ref{fig:plot}(c), we compare the gain achieved for two registration window durations, \(T_{\mathrm{reg}} = 250~\mu\text{s} \text{ and }400~\mu\text{s}\), while keeping \(D_b = 150~\mu\text{s}\) and \(R_C = 614.4\)~Mbps fixed. The results show that increasing \(T_{\mathrm{reg}}\) results in a gradual reduction in the gain achieved by the proposed framework. This is because a larger registration window leads to greater backlog during registration cycles, thereby increasing delay. In contrast, the dedicated-wavelength baseline remains unaffected, as registration is isolated from data-carrying wavelengths. 

\section{Conclusion}
\label{sec:conclusion}

This paper proposed a scheduling framework for TWDM-EPON-based fronthaul networks that facilitates periodic RU registration without the spectral inefficiency of a dedicated registration wavelength. By hosting the registration process on any selected wavelength and temporarily redistributing active ONUs to the remaining ones, the framework accommodates the registration window while meeting the latency and jitter bounds required for fronthaul traffic. 

Numerical evaluations demonstrate that this approach significantly outperforms the ITU-T baseline of wavelength dedication. Specifically, the proposed method achieves up to a 71\% increase in supported RUs per OLT, markedly improving system capacity and resource efficiency. These results confirm that the proposed traffic redistribution strategy to bypass "quiet zones" is an effective strategy for meeting the stringent demands of fronthaul traffic.

% \section*{Biography Section}
 

% \end{IEEEbiography}



\vfill

\end{document}